\documentstyle[prl,twocolumn,aps,epsf]{revtex}
\begin{document}
\twocolumn[\hsize\textwidth\columnwidth\hsize\csname@twocolumnfalse%
\endcsname
\title{Dynamics of Holes and Universality Class of the Antiferromagnetic 
Transition in the Two Dimensional Hubbard Model}
\author{F. Guinea$^{\dag}$, E. Louis$^{\ddag}$,
M. P. L\'opez-Sancho$^{\dag}$ and J. A. Verg\'es$^{\dag}$}
\address{$^{\dag}${I}nstituto de Ciencia de Materiales de Madrid,
CSIC, Cantoblanco, E-28049 Madrid, Spain.
$^{\ddag}$ Departamento de F{\'\i}sica Aplicada,
Universidad de Alicante, Apartado 99, E-03080 Alicante, Spain.}
\date{\today}
\maketitle
\begin{abstract}
The dynamics of a single hole (or electron) in the
two dimensional Hubbard model
is  investigated. The antiferromagnetic background is
described by a N\`eel state, and the hopping of the carrier
is analyzed within a configuration interaction approach. 
Results are in agreement with
other methods and with experimental data when available. 
All data are compatible with the opening  
of a mean field gap in a Fermi liquid of spin polarons, the so called Slater
type  of transition. In particular, this hypothesis explains
the unusual dispersion relation of the quasiparticle bands near
the transition. Recent photoemission data for Ca$_2$CuO$_2$Cl$_2$ are
analyzed within this context.
\end{abstract}
]
\narrowtext
The nature of the insulating phase of the Hubbard model 
in a bipartite (square) lattice has attracted
a great deal of attention. A clarification of the universality class
of the transition will greatly help in understanding the nature of the 
^^ ^^ metallic ", gapless, phase away from half filling.
A detailed discussion of the issue can be found in~\cite{IFT98}.
Opposing viewpoints on the problem are presented in~\cite{La97,AB97}.
An appealing approach is the scaling analysis discussed in~\cite{IFT98}.
It is assumed that physical quantities can be expressed
in terms of a single energy scale, $\Delta$, which vanishes at the
transition. A wealth of numerical results suggest that such
scaling does exist. The data implies that typical length
scales, $\xi$, scale as $\Delta^{\frac{1}{4}}$. This implies
the existence of a quasiparticle band, in the gapless side,
with a dispersion relation $\epsilon_k \propto
k^4$\cite{IA98,AI98}. 


The understanding of the Mott transition as function of filling
in the Hubbard model in two dimensions (2D) requires a knwoledge of
the dynamics of carriers introduced into the 
undoped antiferromagnetic (AF) system. This problem has been 
studied in the related $t$--$J$ model\cite{DN94,LG95}.
The bands obtained by numerical methods are well described
within the Born approximation\cite{MH91},
which accounts for hopping in one sublattice only.
It has been shown that
higher order corrections are small\cite{LM95,EJ95}.
The bandwidth scales, approximately,
with $J = 4 t^2 / U$.
This can be understood from the fact that a spin flip
is required to restore the AF background 
to its original texture after a hole hops between equivalent 
sites. Numerical results for the Hubbard model\cite{BS94,PL95},
show similar features, with a narrow band which is very flat
between the $( 0 , \pi )$ and $( \pi / 2 , \pi , 2 )$ points of the
Brillouin Zone. This band is also reasonably described by
a generalization of the Born approximation to the
Hubbard model\cite{ABetal95}. As for the $t$--$J$ model, this picture
suggest that numerical results are well described 
in terms of dressed holes which hop within a given sublattice.

The same picture is obtained from mean field
calculations. It has been shown that single holes in the
Hubbard model induce the formation of inhomogeneous
spin textures\cite{VL91}, which can be described as
spin polarons, or spin bags \cite{SWZ89}. An improved wavefunction results from 
hybridization of polarons localized at different sites, so
that the translational symmetry is restored\cite{LC93b}.
This is a standard procedure in molecular physics,
where Hartree--Fock (HF) solutions are improved by the
Configuration Interaction (CI) method. The resulting polaron
band is in reasonable agreement with other calculations\cite{LG98,LG98b}.
 
In the following, we analyze the implications of the spin polaron band
of the simple Hubbard model for the nature of the Mott transition. 
We also extend these calculations to include second and third
nearest neighbor hoppings, in order to compare directly to
experiments\cite{Ki98,Ro98}. Our results suggest that
a mean field picture is a reasonable starting point for the
analysis of the dynamics of holes in the Hubbard model
in 2D, providing an intuitive picture 
of the Mott transition, and reproducing
adequately the main features observed in the experiments. 

The usual HF approximation to the Hubbard model
with nearest neighbor couplings only in a square lattice,
gives an AF ground state at half filling.
This solution reproduces the existence of a gap in the
charge excitations, which scales with $U$, for large $U$.
The approximation misses the low energy spin waves, which can
be incorporated by including the
low amplitude fluctuations around the local minimum
defined by the HF solution, that is, the
Random Phase Approximation (RPA). The RPA reproduces the correct
result in the large $U$ limit\cite{SWZ89,SS95}.
Thus, the mean field solution is a good starting point for 
the analysis of the hole dynamics.

\begin{figure}
\begin{center}
\mbox{\epsfxsize 5cm \epsfbox{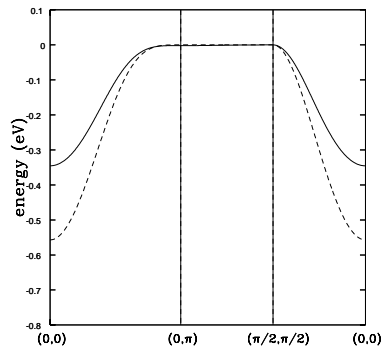}}
\end{center}
\caption{
Quasiparticle band structure for a single hole in the Hubbard
band with $t = 0.35$eV and $U = 3.92$eV. Hartree-Fock approximation:
dashed line. CI approximation (full line): dispersion relation, see Eq. (3),
fitted to the numerical results for a  $14 \times 14$ cluster.}
\label{fig:band0}
\end{figure}

A naive generalization of the AF solution to the doped case can be obtained 
by filling the lowest lying states in the upper Hubbard band.  
Fig. \ref{fig:band0} shows this (HF) band in the extendend Brillouin Zone,
for $t = 0.35$ eV and $U = 3.92$ eV, in order to compare to experiments
(see below). Due to the AF background, the bands have an additional 
periodicity, so that the point $( \pi , \pi )$ is degenerate 
with $( 0 , 0 )$. As the Hubbard model
in the absence of other hoppings has electron-hole symmetry,
the same band describes electrons.
The HF band shows a flat dispersion along the $( 0 , \pi ) \rightarrow
( \pi / 2 , \pi / 2 )$ direction. 
The HF band can be written in the extended zone scheme as:
\begin{equation}
\epsilon_k = \frac{\Delta}{2} - \sqrt{\frac{\Delta^2}{4} +
4 t^2 [ \cos ( k_x ) + \cos ( k_y ) ]^2}  
\label{hf}
\end{equation}
\noindent where $\Delta = \frac{U}{2}\langle n_{\uparrow}-n_{\downarrow}
\rangle$.
Near the upper edge, this band can be approximated by:
$ \epsilon_k \approx ( 4 t^2 / \Delta ) [ \cos ( k_x ) +
\cos ( k_y ) ]^2$. It describes a hole with effective hoppings
$t^2 / \Delta$ to the neighbor in position $( 2 , 0 )$,
and $2 t^2 / \Delta$ to that at $( 1 , 1 )$. Hopping between
different sublattices is not allowed. 
Around the point $( 0 , \pi )$, we can expand:
$\epsilon_k \approx ( t^2 / \Delta ) ( k_x^2 - k_y^2 )^2$.
This quartic dispersion explains well the findings 
reported in\cite{IA98,AI98}. 

The integrated density of states
is dominated by the flatness of the dispersion around
$( 0 , \pi )$. Integrating over $\vec{k}$, we find,
in the Hartree Fock case,
$D ( \omega ) \sim \log [ t^2 / ( U \omega ) ] / \sqrt{\omega}$
near the upper band edge, and a similar dependence for
the polaron band. 

The state $\tilde{c}_{k,\uparrow}^{\dag} | \Psi_0 \rangle$, where
$  | \Psi_0 \rangle$ is the Slater determinant which describes
the AF insulator, has negative
compressibility and is unstable against charge inhomogeneities \cite{VL91}. 
This topic was first discussed in\cite{Vi74}. As the compressibility, $\kappa$,
is proportional to $( \partial \mu / \partial n  )^{-1}$, it
will be negative if the chemical potential  is lowered upon doping 
\cite{AGG98}. The states near the gap edges are almost perfectly
localized in one sublattice or the other. Thus, the additional electrons
reduce the value of $\langle n_\uparrow - n_\downarrow \rangle$ by an
amount directly proportional to the density. The gap is lowered by
$\frac{Un}{2}$, where $n$ is the hole density per lattice site. 
Because of the singularity in the density of states at the gap edge,
the difference between the chemical potential and the gap edge increases
as $\delta \mu \propto n^2$, 
neglecting logarithmic corrections. This dependence cannot
compensate the decrease in the gap, $\propto n$, and the chemical
potential is lowered upon doping.
The transverse spin susceptibility also shows
an instability as a consequence of the singularity in the
density of states. 
Note, however, that the flatness of the band, which is a typical precursor
of ferromagnetism\cite{VB97}, does not imply a ferromagnetic instability.
The interaction between carriers of opposite spin is greatly reduced,
as they are localized in different sublattices.
This analysis is in agreement with general studies of the possibility
of phase separation in the Hubbard model\cite{Vi74}.
Similar results can be rigourously demonstrated in the limit of 
infinite dimensions\cite{Do96}.

The previous shortcoming of the homogeneous mean field solution of
the doped Hubbard model can be overcome by considering generic charge and spin
textures\cite{VL91}. Among the many different solutions which are extrema
of the Hartree Fock solutions, spin polarons and domain walls tend to be
the most stable. We now analyze in detail the spin polaron, which is 
more stable
at large values of $U/t$, and its energy is further lowered by delocalization
effects (see also below). 
The spin polaron, in the large $U$ limit,
can be approximately described as a self localized
electron (or hole) for a cluster which contains at least five sites, and with
a ferromagnetic alignment of the spins in its interior. The overlap between
this solution and homogeneous solutions decays with the size of the cluster
as $1/L^2$\cite{LC93a}, implying the formation of a  
localized state. This state is separated from the upper (lower) Hubbard band
by an energy of order $t$ in the large $U$ limit. This splitting
suppresses the contribution of the extra electrons to the spin
susceptibility, curing the instabilities of the homogeneous
HF approximation.  In addition to this state, another
localized level splits off from the lower (upper) band, and moves to
an energy of order of a fraction of $U$ from that band.
Thus, the low energy spectral
weight per electron (hole) is greater than one,
in qualitative agreement with the arguments given 
in\cite{EM91}. The Slater determinant which describes
a solution of this type can be written as:
\begin{equation} 
| \Psi_i \rangle = 
\tilde{c}^{\dag}_{i,\uparrow} \Pi_i \left( \sum_k \alpha^i_k 
\tilde{c}^{\dag}_{k,s}
\right) \left( \sum_{k'} \beta^i_{k'} \tilde{c}_{k',s} \right) | \Psi_0 \rangle
\label{polaron}
\end{equation} 
where $| \Psi_0 \rangle $ describes the antiferromagnetic
insulator at half filling. Thus the wavefunction $| \Psi_i \rangle$
is obtained from  $| \Psi_0 \rangle $ by
adding a given number of electron-hole excitations, plus an electron (hole)
at site $i$.

As discussed elsewhere\cite{LG98,LG98b}, corrections
to this inhomogeneous
HF solution can be of two types: low amplitude fluctuations,
which can be studied within the RPA\cite{GLV92}, and hybridization
of solutions centered at different lattice sites\cite{LC93b}.
The combination of these solutions leads to wavefunctions of the type
$| \Psi_k \rangle = \sum_i e^{i \vec{k} \vec{r}_i }
| \Psi_i \rangle $, and to the formation of a polaron
band, whose width scales as $t^2/U$ in the limit of large $U$. 

The dispersion  band  for a $14 \times 14$ cluster and the same values of 
$U$ and $t$ used above, is shown in Fig. \ref{fig:band0}. The polaron 
(CI) band has the same shape than the HF band but is substantially narrower 
(around a 40\%). We adscribe this narrowing to the fact that the polaron band
is more strongly dressed by spin excitations (note, however, that
the approximation used here only includes longitudinal, Ising like, modes).
Because of the extension of the individual polarons, the band cannot be 
parametrized
in terms of a few hopping parameters, as in the homogeneous solutions
described previously. The numerical results of Fig. \ref{fig:band0}
can be accurately fitted by,

\begin{eqnarray}
\epsilon_{\bf k}&=& \epsilon_0 + 4 t_{11} \cos (k_x ) \cos ( k_y )
 + 2 t_{20} [ \cos ( 2 k_x ) + \cos ( 2 k_y ) ] \nonumber \\&&
+  4t_{22} \cos ( 2 k_x ) \cos ( 2 k_y ) +
t_{31}[ \cos ( 3 k_x) \cos ( k_y )\nonumber \\&& + \cos ( k_x ) 
\cos ( 3 k_y )] + 2 t_{40} [ \cos ( 4 k_x ) + \cos ( 4 k_y ) ].
\end{eqnarray}

\noindent with $t_{11} = 0.130542$ eV , $t_{20} = 0.062056$ eV,
$t_{22}= -0.006130$ eV, $t_{31}=-0.003963$ eV, and $t_{40}=-0.000836$
eV. In the $(0,\pi-\eta)$ direction we can expand the dispersion
relation for small $\eta$ as,
\begin{eqnarray}
\epsilon_{\bf k}&\approx&\epsilon_0 + 4(-t_{11}+t_{20}+t_{22}-2t_{31}+t_{40}) 
\nonumber \\ && + 2\eta^2(t_{11}-2t_{20}-4t_{22}+10t_{31}-8t_{40})
\nonumber \\ && + \frac{\eta^4}{3}(-\frac{1}{2}t_{11}+4t_{20}+8t_{22}-
41t_{31}+64t_{40})\;.
\end{eqnarray} 
\noindent Using the parameters given above we obtain $\epsilon_{\bf k}\approx
-0.004\eta^2+0.08\eta^4$. This result shows that the
quartic term dominates, as in HF and in agreement with \cite{IA98,AI98}.

\begin{figure}
\begin{center}
\mbox{\epsfxsize 5cm \epsfbox{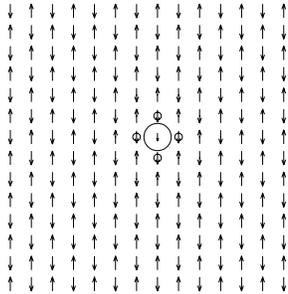}}
\end{center}
\caption{Spin (arrows) and charge distribution around a
localized hole in a $14 \times 14$ cluster
using the parameters from Ref. [20].}
\label{fig:texture}
\end{figure}

Experiments on the dynamics of holes in CuO planes suggest that a Hubbard
model with nearest neighbor hoppings only is 
insufficient\cite{EO97,Ki98,Ro98}. The inclusion of additional hoppings
is straightforward within the scheme discussed here. 
We have calculated
the electron and hole bands using the parameters given in\cite{Ro98}:
$U = 3.92$ eV, $t = 0.35$eV ($J$=0.125 eV), $t' = -0.12$eV and $t'' = 0.08$eV.
For hole doping this choice of parameters leads to the formation of
well localized spin polarons at the mean field level. 
The spin and charge textures associated to the (hole)
spin polaron are shown in [\ref{fig:texture}], for a $14 \times 14$
cluster.
We find no inhomogeneous solutions for electron doping (see below)

The quasiparticle bands are shown in fig.[\ref{fig:band}].
The new hoppings break the electron--hole 
symmetry present in the simple Hubbard model and induce
a quadratic dispersion at the band edges, which dominates the
quartic terms discussed so far. This dispersion relation implies that,
upon doping, emptying the lowest lying states give 
a change in the chemical potential $\delta\mu\propto n$ that can 
be compensated by the decrease in the gap $\propto n$. As a result,
homogeneous mean field solutions can be stable.
The HF and CI bands
have a similar shape, while, as in the case of the simple Hubbard model, the CI
band is much narrower than the CI band (an approximate reduction of 40\% is
noted). As remarked
in \cite{Ro98} the HF band is much wider than the one given by 
photoemission experiments. The CI band is in much closer agreement with
the observed band (approximately 0.4 eV wide) and with the numerical 
results for the corresponding $t$--$J$ model reported in \cite{Ki98}.
The shape of the hole band shown in Fig. 2 is also in full
agreement with experiments. 
The CI band can  be fitted with the dispersion relation of Eq. (3).
In this case the parameters are $t_{11}=-0.071428$ eV, 
$t_{20}=  0.195491$ eV, $t_{22}=-0.016190$ eV, $t_{31}=0.003486$ eV, 
and $t_{40}=-0.006278$ eV. The expansion near (0,$\pi$) gives
${\rm constant}-0.63\eta^2+0.05\eta^4$, which is mainly quadratic.

\begin{figure}
\begin{center}
\mbox{\epsfxsize 5cm \epsfbox{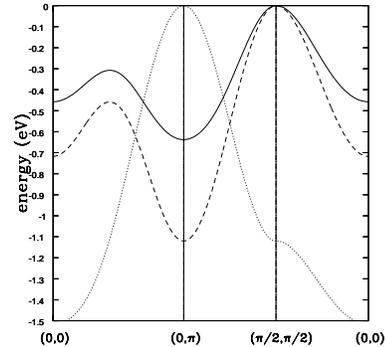}}
\end{center}
\caption{
Quasiparticle band structure for a singly doped
Hubbard model with second and third nearest neighbor hoppings,
using the parameters from Ref. [20]
Hartree-Fock approximation:
dashed line (hole doping) and dotted line (electron doping). 
CI approximation (full line): dispersion relation, see Eq. (3), fitted to 
the numerical results for a $14 \times 14$ cluster.}
\label{fig:band}
\end{figure}
 
A detailed analysis of the hole dispersion
relation in the corresponding $t-t'-t''-J$  model is presented
in\cite{TS99}. 
It is instructive to compare our results with those there.
As mentioned at the beginning, the nature of the approximation used here
for the Hubbard model describes similar processes to those included in
the Self Consistent Born Approximation used in\cite{TS99}. However, when
reducing the Hubbard hamiltonian to the $t-J$ one, hopping terms
$\propto J$ are neglected. Thus, the bandwidth obtained from the
$t-J$ hamiltonian should be smaller than the exact.
The approximation used here
neglects the interaction of the hole with transverse spin waves,
which probably leads to an overestimation of the exact bandwidth.
Magnetic correlations around the hole are in qualitative agreement
with\cite{TS99} (see fig.[\ref{fig:texture}]).

For electron doping, instead, the homogeneous solution is stable. The
HF band is shown in Fig. [\ref{fig:band}]. 
The band has a significantly smaller effective
mass near (0,$\pi$) than in the case of hole doping, in
agreement with \cite{Ki98}, resulting in a larger quasiparticle
peak as observed experimentally \cite{Ki98}.  

The instability of the homogeneous mean field solution for
the doped Hubbard model can also lead to the formation of stripes,
or domain walls \cite{VL91,ZG89,PR89,Sc90}. Similar solutions have been
found in the $t$--$J$ model\cite{WS97,HM98}).
A transition from the undoped insulator to a doped system
with stripes should be qualitatively different from the one described above.
The stripes will remain, most likely, static, even after the inclusion of
corrections beyond HF. The lack of long range magnetic order
can be understood from the weakening of the effective exchange coupling,
and the enhancement of fluctuations\cite{CH96}, as for the spin
polaron solutions.
Thus, we cannot rule out the possibility of
a sharp discountinuity, as function of $U/t$ and for a fixed low doping,
between a gapless phase with delocalized spin polarons, and another
with static domain walls.

The above analysis shows that the properties of the
insulating and the lightly
doped phases of the Hubbard model can be reasonably understood within
straightforward extensions of
mean field theory. The picture is similar to the
spin bag model \cite{SWZ89,KS90a,KS90b}.
The main difference is that the spin bag approach uses homogeneous
solutions as starting point. We have addressed
the instabilities of these solutions, and showed how they can be 
overcome. The adequacy of mean field theory in describing the 
metal-insulator transition is due to the suppression of dynamic
charge fluctuations by the formation of the staggered magnetization,
as in infinite dimensions\cite{CK98}.
The resulting picture leads to a flat quasiparticle band, in agreement
with detailed numerical calculations. The density of states is singular
at the band edges, enhancing the effects of the interactions. A quartic
dispersion relation is obtained around the point (0,$\pi$).
Including second and third nearest neighbors hoppings allows us to compare 
with recent photoemission data. This new hoppings enhance the
quadratic terms in the dispersion relation and break the
electron/hole symmetry. The CI hole band is much narrower than
the HF band and in agreement (both shape and width)
with the experimental results.

Financial support from the  CICYT, Spain, through grants PB96-0875,
PB96-0085, PB95-0069, and from CAM, Madrid, Spain, 
is gratefully acknowledged.

\end{document}